\documentclass[runningheads]{llncs}

\usepackage{eccv}

\usepackage{eccvabbrv}

\usepackage{graphicx}
\usepackage{booktabs}

\usepackage[accsupp]{axessibility}  

\usepackage{graphicx}
\usepackage{subcaption}
\usepackage{amsmath}
\usepackage{graphicx} 
\usepackage{amssymb}   
\usepackage{amsmath}   
\usepackage{xcolor}
\usepackage{colortbl}
\usepackage{booktabs}
\usepackage{graphicx}
\usepackage{booktabs}  
\usepackage{multirow}  
\usepackage{graphicx}  
\usepackage[table,xcdraw]{xcolor} 
\usepackage{float}
\usepackage[utf8]{inputenc}
\usepackage{graphicx} 
\usepackage{subcaption} 
\usepackage{amssymb}  
\usepackage{pifont}   
\usepackage{graphicx} 
\usepackage{booktabs} 
\usepackage[table]{xcolor} 
\usepackage{bbding}

\usepackage{hyperref}

\usepackage{orcidlink}

\begin{document}

\title{A First Exploration of Neuromorphic OT-CFM for Multi-Speaker VSR} 


\author{Lin Chen\inst{1}\orcidlink{0009-0009-8696-1788} \and Jingping Fang \inst{1}\orcidlink{0009-0006-8300-6599} \and Hairui Liu\inst{1}\orcidlink{0009-0008-8687-7642} \and Chenyang Xu\inst{2}\orcidlink{0009-0000-7625-3117} \and \\Junhao Chen\inst{3}\orcidlink{0009-0001-5916-9789} \and Xiaorui Li\inst{1}\orcidlink{0009-0008-5225-0956} \and Weidong Cai\inst{4}\orcidlink{0000-0003-3706-8896} \and Xiaoming Chen\textsuperscript{\Envelope}\inst{1}\orcidlink{0000-0002-7503-3021}}

{\renewcommand{\thefootnote}{}\footnotetext{\textsuperscript{\Envelope}Corresponding author.}}


\authorrunning{Lin Chen et al.}

\institute{Beijing Technology and Business University, China \and Xidian University, China \and Tsinghua University, China \and The University of Sydney, Australia \\ \email{Corresponding Author: Xiaoming Chen (xiaoming.chen@btbu.edu.cn)}}

\maketitle

\begin{abstract}
Visual Speech Recognition (VSR) tasks in complex multi-speaker scenarios are severely hindered by rapid head motions, occlusions, and subtle lip articulations. Traditional RGB-based methods struggle here due to low rates and motion blur of frames. To overcome these, we propose \textbf{LipsFlow}, a neuromorphic-inspired VSR framework that converts RGB videos into high-temporal-resolution event streams. For multi-speaker, we employ ByteTrack tracking and TalkNet active speaker detection to temporally segment scenes into single-speaker clips, enabling focused per-speaker analysis. By explicitly capturing microsecond-level articulatory dynamics via learnable event-based representations, LipsFlow achieves inherent robustness against visual degradation. To efficiently model these dense event-based features and adapt to speaker-specific articulatory patterns, we introduce Optimal Transport Conditional Flow Matching (\textbf{OT-CFM}). It enforces deterministic, straight-line trajectory generation in a semantic latent space, slashing inference latency to just two Ordinary Differential Equation (ODE)  steps. Furthermore, we design a Dual-Level Semantic Supervision mechanism combining token-level BERT weight tying and sentence-level priors to resolve homophene ambiguities. Validated on competitive benchmarks, LipsFlow achieves a state-of-the-art \textbf{WER of 22.3\% at 240 ms latency}, establishing a highly robust and efficient paradigm for event-based VSR.
\end{abstract}

\section{Introduction}
\label{sec:intro}

VSR advances have driven the evolution of multimodal interaction, bridging the sensory gap when traditional acoustic signals are compromised. However, in complex multiple-speaker scenarios, establishing robust visual models independent of auditory cues is significant for sustained communication \cite{Shi22-AVHuBERT, Ma23-AutoAVSR,Haliassos23-RAVEn,Zhu23-VATLM}. While current mainstream research \cite{Ma23-AutoAVSR,Zhu23-VATLM,Haliassos23-RAVEn,Hao25-LipGen,bulzomi2023end} typically relies on standard active pixel sensors, the inherent limitations of frame-based mechanisms, specifically low fixed frame rates and motion blur, result in the loss of transient visual cues, such as the rapid lip closures of plosives (e.g., /p/ and /b/) that occur within milliseconds~\cite{Gallego22-EventSurvey}. To mitigate these sensing limitations, recent works \cite{Tan22-MSTP,bulzomi2023end} have introduced neuromorphic event streams to the VSR task, demonstrating the potential of asynchronous event streams to capture microsecond-level lip dynamics~\cite{Tan22-MSTP}. Furthermore, Arriandiaga \textit{et al.}~\cite{Arriandiaga19-AVT} have explored event-driven cameras for target speaker enhancement in complex multi-speaker scenarios.

Despite these advancements, current VSR paradigms with event streams \cite{Tan22-MSTP,Arriandiaga19-AVT,bulzomi2023end} face substantial bottlenecks in both representation and generation. 
First, in terms of vision representation, traditional methods \cite{Pan24-LLMVSR,Liu2023SynthVSR} typically aggregate asynchronous events into dense frames, feeding them into standard backbones designed for synchronous data. This approach often fails to disentangle fine-grained phonetic movements from non-phonetic oral noise (e.g., chewing or jitter). 
Second, the text sequence generation aspect remains a critical efficiency limitation. Dominant approaches rely either on Auto-Regressive (AR)~\cite{Ma23-AutoAVSR}, which suffer from prohibitive token-by-token latency, or on standard diffusion models~\cite{Liu2023SynthVSR}, which require extensive stochastic sampling steps to converge, thus affecting text decoding efficiency. Finally, these feed-forward architectures often lack the iterative refinement capabilities needed to retroactively resolve visual ambiguities (e.g., distinguishing ``I love you baby'' from ``Elephant juice bat''). 

\begin{figure*}[t]
  \centering
  \begin{subfigure}[b]{0.48\textwidth}
    \centering
    \includegraphics[width=\linewidth]{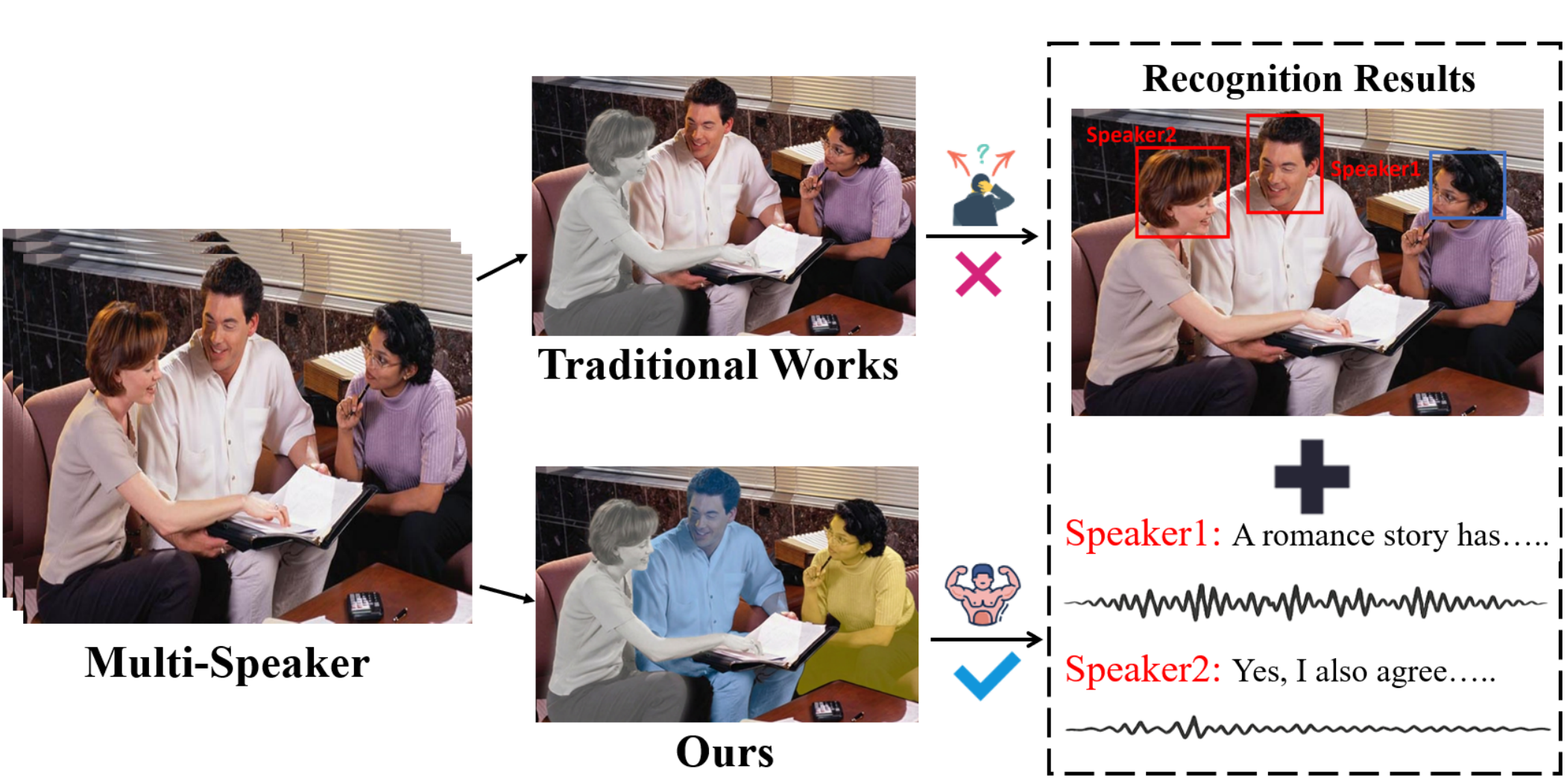}
    \caption{Comparison of traditional works and ours with multi-speaker recognition.}
    \label{sss}
  \end{subfigure}
  \hfill
  \begin{subfigure}[b]{0.48\textwidth}
    \centering
    \includegraphics[width=\linewidth]{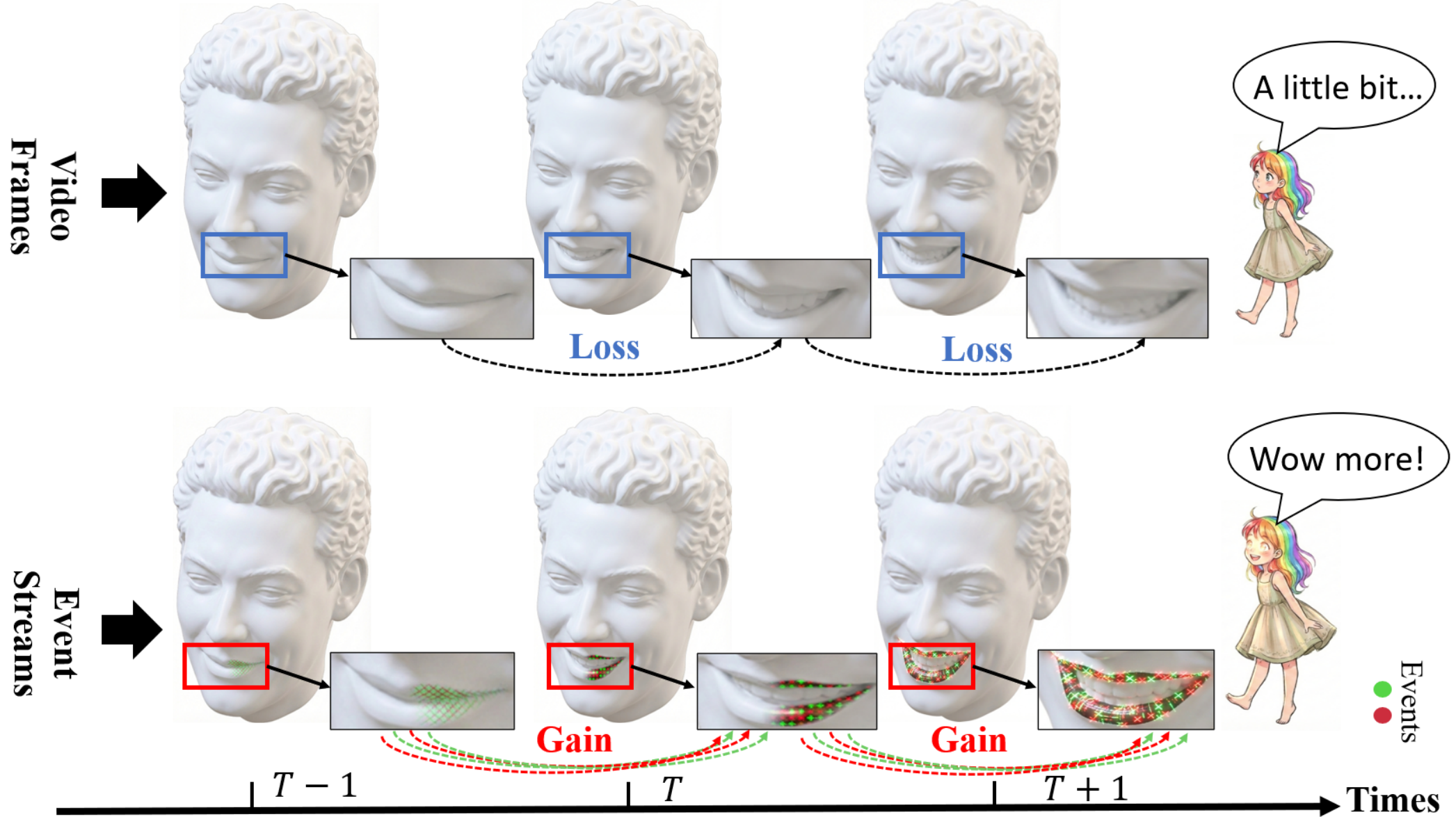}
    \caption{Video frames and event streams in terms of visual information loss and gain.}
    \label{fig:right_results}
  \end{subfigure}

  \caption{Our work simultaneously focuses on the following two challenges: \textbf{(a)} Traditional VSR methods are only applicable to single-speaker scenarios, while we are able to adapt to multi-speaker scenarios; \textbf{(b)} Compared with video frames, event streams provide more subtle information gain for lip reading, thus compensating for the loss of fine-grained motion between consecutive frames.}
  \label{fig1}

\end{figure*}

To overcome these limitations, we simultaneously tackle the challenges of \textit{Multi-Speaker}\textbf{ }and \textit{Frame Loss}, as illustrated in Fig. \ref{fig1}. Reconstructing transient phonetic micro-dynamics and resolving the inherent visual ambiguity of lip reading naturally formulate VSR as a complex sequence generation task.
While inspired by the powerful diffusion modeling \cite{liu2024instaflow,Mehta24-Matcha,Choi25-V2SFlow,liu2022flow,Liu2023SynthVSR}, we explore its potential for VSR applications. We further introduce OT-CFM decoder based on recent research of Flow Matching \cite{Lipman23-FlowMatching,Guo24-VoiceFlow,liu2024instaflow,Kornilov24-OptimalFlow,liu2022flow} to the VSR domain for the first time. By extending this technique to multi-speaker scenarios and explicitly learning straight trajectories that minimize transport costs, our method achieves a superior trade-off between recognition fidelity and inference efficiency, significantly accelerating model speed without compromising accuracy.

In this work, we propose \textbf{LipsFlow}, an end-to-end neuromorphic event-based VSR framework explicitly tailored for complex multi-speaker scenarios. Our method employs a specially designed Event Representation Module for lip reading within multi-speaker environments, explicitly separating fine-grained speech dynamics from non-speech noise, such as chewing or temporal visual occlusion. To resolve homophene ambiguities and ensure global linguistic coherence, we establish a Dual-Level Semantic Supervision strategy leveraging BERT weight tying and Sentence-BERT priors. By combining this strategy with OT-CFM, we explicitly model the generation of continuous, optimal transmission paths for VSR, thereby achieving faster and more accurate training and inference, effectively addressing the fundamental limitations inherent in purely visual methods.

\section{Related Works}

\textbf{Visual Speech Recognition Paradigms.} VSR has become indispensable for robust speech decoding in acoustically degraded environments \cite{Ma23-AutoAVSR, Zhu23-VATLM}. 
Early end-to-end pioneers \cite{Assael2016LipNet} leveraged 3D-CNNs and Connectionist Temporal Classification (CTC) loss to bypass complex frame-level alignment. 
Recent advancements \cite{Shi22-AVHuBERT,Ma23-AutoAVSR} have pushed performance boundaries through massive self-supervised pre-training, which model long-range dependencies via Conformer architectures. Furthermore, state-of-the-art (SOTA) approaches \cite{Hao25-LipGen,Pan24-LLMVSR,Shi24-CMVSR} have attempted to enhance robustness via synthetic visual data generation, large language model priors or multi-modal contrastive learning. 
However, above these systems face an insurmountable physical bottleneck: reliance on standard RGB cameras. Operating at 25--30 FPS, the camera suffers from severe motion blur and inevitably undersamples  rapid articulatory micro-dynamics.

\noindent\textbf{Flow Matching Modeling. }To address the text mapping in VSR caused by homophenes, generative modeling has evolved from AR sequence prediction to Non-Autoregressive (NAR) backbones. 
While diffusion-based models \cite{Peebles23-DiT,liu2024instaflow,Liu2023SynthVSR} improve generation fidelity by iteratively refining noise, they necessitate 50--1000 sampling steps, incurring prohibitive latency and computational costs that hinder real-time deployment. 
Conditional Flow Matching (CFM) \cite{Lipman23-FlowMatching} has recently emerged as a superior generative backbone. Compared with the diffusion, which models curved stochastic trajectories, CFM with Optimal Transport (OT) objectives regresses straight-line probability paths \cite{liu2022flow}, enabling the ODE solver \cite{Chen18-NeuralODE} to traverse from prior noise to target data in 2-10 NFE steps. 
While OT-CFM \cite{Kornilov24-OptimalFlow} has revolutionized speech generation works \cite{Mehta24-Matcha,Guo24-VoiceFlow,Liu24-SpeechFlow,ren2021fastspeech}, its potential in VSR remains unexplored, particularly  in multi-modal contrastive learning \cite{Choi25-V2SFlow}. By combining efficient OT-CFM\cite{Kornilov24-OptimalFlow} generation with Dual-Level Semantic Supervision \cite{Reimers19-SBERT}, our approach not only resolves semantic ambiguities inherent to lip reading but also achieves real-time inference speed.

\noindent\textbf{Event-based Representations. }Neuromorphic event cameras offer an inspiration for VSR by capturing asynchronous brightness changes with microsecond-level temporal resolution, high dynamic range ($>120$ dB), and immunity to motion blur \cite{Gallego22-EventSurvey}. These properties have driven success in high-speed tasks such as optical flow estimation, object tracking, and detection using recurrent transformers \cite{Gehrig23-RVT}. Despite these sensory advantages, considering event cameras in  VSR presents unique challenges due to the sparse nature of events in textureless lip reading and the difficulty of representation learning. Existing event-based VSR methods \cite{Zhang25-MTGA,Dampfhoffer24-SpikGRU,Tan22-MSTP} often aggregate asynchronous events into dense representations like Voxel Grids or Event Stacks to utilize standard CNN backbones. However, these methods typically struggle in unconstrained multi-speaker environments. Generic event-stream encoders fail to disentangle target phonetic movements from non-communicative oral activities (e.g., chewing, yawning) or background noise, as both generate dense event streams. Furthermore, unlike RGB-based VSR which benefits from massive datasets \cite{Ma23-AutoAVSR}, event-based VSR suffers from data scarcity. While simulation techniques Video-to-Events (V2E) \cite{Gehrig20-Video2Events} exist, there is a lack of frameworks that effectively bridge the gap between large-scale RGB pre-training and high-frequency event sensing.

\section{Proposed Method}
\label{framework}
\subsection{Overview}

\begin{figure*}[t]
    \centering
    \includegraphics[width=1\linewidth]{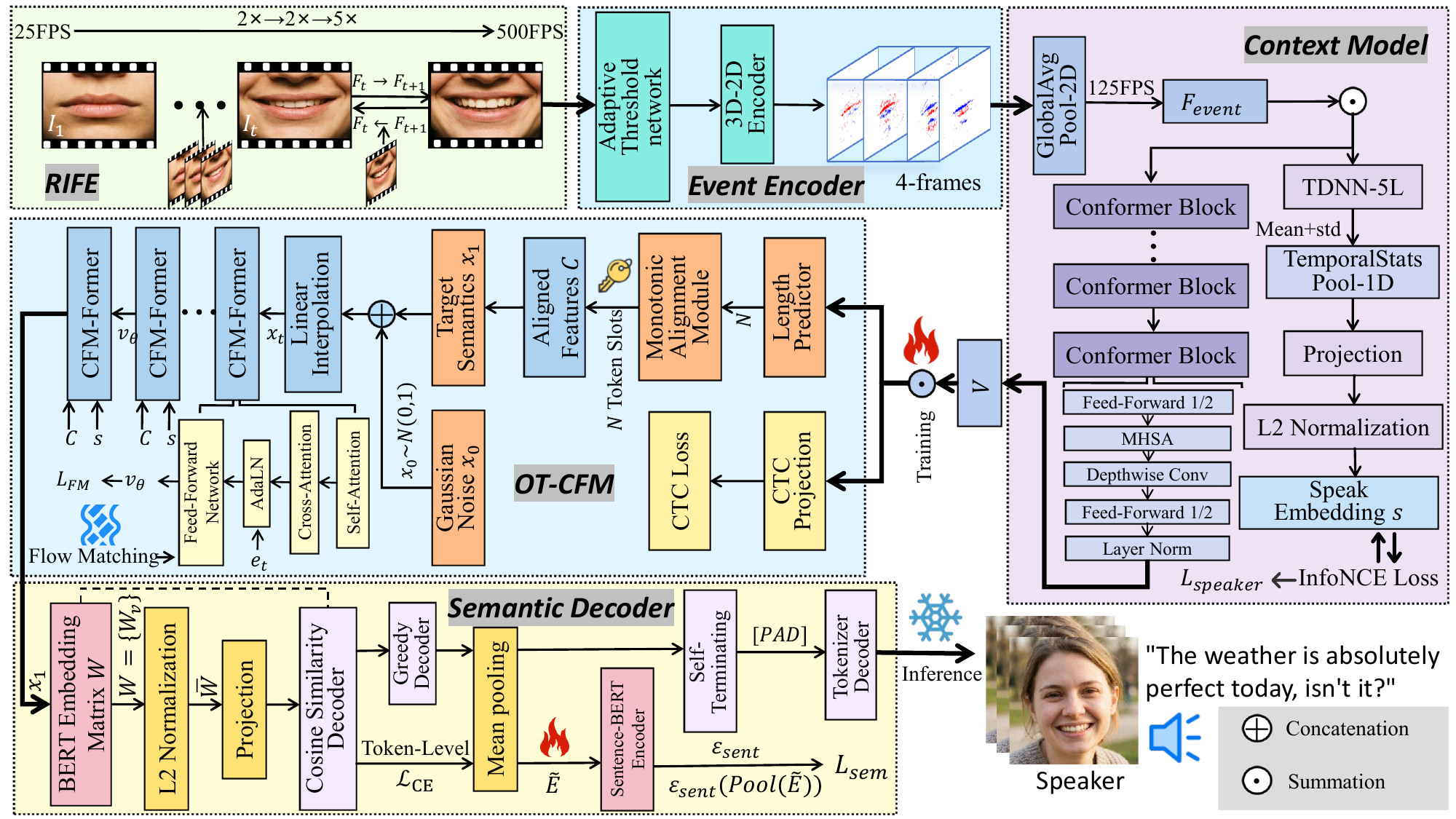}
    \caption{\textbf{Overview of LipsFlow Framework.} Event-based visual features are extracted via hierarchical interpolation and Conformer encoding, then conditioned on speaker embeddings through OT-CFM to generate semantic token representations. The BERT \cite{Devlin19-BERT} weight tying is utilized alongside multi-level supervision signals during training (token-level $\mathcal{L}_{\mathrm{CE}}$, sentence-level $\mathcal{L}_{\mathrm{sem}}$, and auxiliary CTC loss $\mathcal{L}_{\mathrm{CTC}}$ for alignment), ensuring both local precision and global semantic coherence.}
    \label{method}
\end{figure*}

We introduce \textbf{LipsFlow} framework in Fig. \ref{method} for handling multi-speaker scenarios in VSR under various complex conditions. In order to tackle multi-speaker scenarios in batches, we employ the ByteTrack~\cite{Zhang22-ByteTrack} front-end to physically isolate each speaker for focused processing.
Rapid articulatory dynamics are captured via a \textbf{Learnable Event-based Representation} (see Sec. \ref{learanbale}) module that extracts high-frequency features from RGB videos.
These features condition our \textbf{Speaker-Conditioned OT-CFM} (see Sec. \ref{speaker-otcfm}), which efficiently generates semantic tokens via a straight-line probability flow. We resolve homophene ambiguities through a \textbf{Prior-Guided Semantic Decoder} (see Sec. \ref{semantic_decoder}), which leverages linguistic priors to enforce global coherence and contextual consistency. 

\subsection{Learnable Event-based Representation}
\label{learanbale}

\noindent\textbf{Hierarchical Frame Interpolation.} Reconstructing high-frequency lip dynamics from RGB videos is an ill-posed problem.
Direct $20\times$ upsampling often fails as large non-linear displacements between frames exceed the effective receptive field of  multi-speaker optical streams, leading to severe ghosting artifacts.
To address this, we devise a hierarchical strategy that decomposes the task into a cascade of manageable stages: $2\times \to 2\times \to 5\times$.
At each level, we leverage a lightweight variant of RIFE~\cite{Huang22-RIFE} to estimate bidirectional optical streams.
By progressively subdividing the temporal intervals, this strategy reduces the motion magnitude between adjacent frames, ensuring robust streams estimation for rapid articulatory motions via the lightweight backbone design.

\noindent\textbf{Adaptive Event Streams Generation.} 
We transform the upsampled frames into high-dynamic-range event streams using a differentiable quantizer \cite{esser2020learned}.
We introduce an adaptive thresholding network that leverages a lightweight CNN to regress a dense  threshold map $\Theta$. An event is triggered at coordinate $\mathbf{x}$ if the logarithmic temporal residual exceeds the local threshold $\theta(\mathbf{x})$:
\begin{equation}
    |\Delta L(\mathbf{x})| = \left| \log I_{t+\Delta t}(\mathbf{x}) - \log I_t(\mathbf{x}) \right| > \theta(\mathbf{x}).
    \label{eq:event_trigger}
\end{equation}
The soft relaxation strategy \cite{Jang17-Gumbel} is employed during training to ensure end-to-end differentiability, enabling the network to selectively amplify articulatory micro-kinematics to better capture subtle and rapid information about lip movements.

\noindent\textbf{Event Encoding and Context Modeling.} To encode the high FPS of event streams, a hybrid 3D-2D encoder (3D Stem + 2D ResNet \cite{He16-ResNet}) aggregates local motion, yielding a feature sequence $F_{\text{event}} \in \mathbb{R}^{T' \times 512}$ at $125$ FPS.
Since $F_{\text{event}}$ captures only local cues, we feed it into the $12$-layer Conformer \cite{Gulati20-Conformer} encoder to model global temporal dependencies.
Synergizing Depthwise Convolution \cite{howard2017mobilenets} and Multi-Head Self-Attention (MHSA) \cite{Vaswani17-Transformer}, the Conformer produces context-aware visual features $V \in \mathbb{R}^{T' \times 512}$. This hierarchical Visual Encoder design allows LipsFlow to effectively model high-frequency transient dynamics of rapid lip movements while maintaining the long-range temporal dependencies.

\subsection{Speaker-Conditioned  OT-CFM}
\label{speaker-otcfm}

\noindent\textbf{Length Prediction and Alignment. }A critical challenge in NAR speech generation is determining the target sequence length $N$ during inference.
Instead of relying solely on implicit attention, we employ a Hybrid Length Predictor \cite{ren2021fastspeech} (combining regression and classification heads) to estimate the number of semantic tokens from the visual features $V$.
Subsequently, a monotonic alignment module projects the high-frequency visual streams ($125$ Hz) onto the predicted $N$ token slots, producing aligned visual conditions $C \in \mathbb{R}^{N \times 512}$.
While the CTC \cite{Graves06-CTC} head is used as an auxiliary loss during training to enforce alignment constraints, the explicit length predictor governs the generation structure during inference.

\noindent\textbf{Speaker Identity Encoding and Disentanglement.}
To disentangle target speech dynamics from multi-speaker interferences, we extract a global speaker embedding $s \in \mathbb{R}^{256}$ using a Time-Delay Neural Network (TDNN) \cite{Waibel89-TDNN} on $F_{\text{event}}$. A temporal statistics pooling layer aggregates identity-specific attributes across time. Crucially, we utilize an InfoNCE \cite{oord2018representation} contrastive loss to enforce neural disentanglement: by clustering same-speaker embeddings, $s$ functions as an active semantic anchor that filters out identity-irrelevant visual noise (e.g., background motion), transcending simple physical cropping.

\noindent\textbf{Optimal Transport Flow Matching.} Rather than regressing discrete tokens, we reformulate generation as learning a continuous probability flow ODE \cite{Chen18-NeuralODE} within a pre-trained semantic latent space.
Building on CFM~\cite{Lipman23-FlowMatching}, we model the transformation from a Gaussian prior $x_0\sim\mathcal{\text{N}}(0,1)$ to BERT-encoded \cite{Devlin19-BERT} target semantics $x_1$. To guarantee real-time efficiency, we adopt the Optimal Transport (OT) path~\cite{Kornilov24-OptimalFlow,liu2022flow,Choi25-V2SFlow}, simplifying our training objective to regressing a vector field that matches the straight-line displacement between these distributions. Moreover, we directly condition this vector field on our aligned visual features $V$ and speaker embedding $s$, ensuring the sequence generation is precisely driven by target lip dynamics and identity. Unlike CTC-based methods\cite{Ma23-AutoAVSR,Assael2016LipNet,Graves06-CTC}, OT-CFM models the distribution of phonetic trajectories, enabling the decoder to resolve visual ambiguities through iterative refinement in the latent space.

\noindent\textbf{Dual-Conditioned Architecture.} We parameterize the velocity estimator $v_\theta$ using a $16$-layer CFM-Former designed to disentangle linguistic content from speaker identity.
At each layer, aligned visual conditions $C$ are injected via Cross-Attention \cite{Vaswani17-Transformer}, where the evolving state $x_t$ serves as queries and $C$ provides keys and values.
Simultaneously, speaker identity is strictly integrated through Adaptive Layer Normalization (AdaLN)~\cite{Peebles23-DiT}:
\begin{equation}
    \mathrm{AdaLN}(h, e_t, s) = \gamma(e_t, s) \cdot \mathrm{LN}(h) + \beta(e_t, s),
    \label{eq:adaln}
\end{equation}
where the affine parameters $\gamma, \beta$ are dynamically modulated by the time embedding $e_t$ and speaker embedding $s$.
This design effectively decouples content generation from identity modulation: Cross-Attention enforces semantic alignment with lip movements, while AdaLN \cite{Peebles23-DiT} injects global speaker traits. Finally, leveraging the geometrically straightened flow path \cite{liu2022flow} modeled by this architecture, we conduct inference by solving the probability flow ODE \cite{Chen18-NeuralODE} via a deterministic $2$-step Euler solver ($\Delta t = 0.5$). This process recovers the target semantic representations $x_1 \in \mathbb{R}^{N \times 512}$.

\subsection{Prior-Guided Semantic Decoder}
\label{semantic_decoder}

\noindent\textbf{Cross-Modal Weight Tying with BERT.} 
To bridge the modality gap between continuous visual kinematics and discrete linguistic semantics, we avoid learning an isolated projection head. Instead, we map the recovered features $x_1$ into the pre-trained BERT~\cite{Devlin19-BERT} embedding space $W$, leveraging it as a strong semantic anchor. Specifically, a learnable projection $\phi(\cdot)$ transforms $x_1$ to align with the semantic manifold of $W$. We then compute the probability distribution over the vocabulary, $\hat{\mathbf{y}}$, via temperature-scaled cosine similarity:
\begin{equation}
\hat{\mathbf{y}} = \mathrm{Softmax}\left( \frac{\phi(x_1) \bar{W}^\top}{\tau \cdot \lVert \phi(x_1) \rVert_2} \right),
\label{eq:cosine_decoding}
\end{equation}
where $\bar{W}$ is the normalized embedding matrix and $\tau$ is a learnable scalar. This formulation enforces directional alignment within the hypersphere, transferring rich linguistic priors to resolve visual ambiguities without massive parameters.

\noindent\textbf{Dual-Level Semantic Supervision.} Although $\mathcal{L}_{\mathrm{CE}}$ ensures local token precision, it is inherently myopic and suffers from lexical rigidity, failing to resolve visual ambiguities arising from homophenes. Therefore, we augment the training objective with a global semantic consistency constraint via a frozen Sentence-BERT~\cite{Reimers19-SBERT} encoder $\mathcal{E}_{\mathrm{sent}}$. To preserve end-to-end differentiability, we bypass the non-differentiable discrete decoding step (e.g., argmax). Instead, we introduce soft-weighted \cite{Jang17-Gumbel} token embeddings as the expectation over the predicted distribution. Specifically, the continuous embedding at step $i$ is calculated as the weighted sum of all fixed BERT \cite{Devlin19-BERT} vocabulary embeddings, where each token's embedding is scaled by its corresponding predicted probability $\hat{y}_{i,v}$ at that step. 

The resulting sequence $\tilde{\mathbf{E}} = [\tilde{\mathbf{e}}_1, \ldots, \tilde{\mathbf{e}}_N]$ is aggregated via mean pooling and fed to Sentence-BERT to minimize the semantic discrepancy:
\begin{equation} 
\mathcal{L}_{\mathrm{sem}} = \left\lVert \mathcal{E}_{\mathrm{sent}}(\mathrm{Pool}(\tilde{\mathbf{E}})) - \mathcal{E}_{\mathrm{sent}}(\mathbf{y}) \right\rVert_2^2 .
\label{eq:semantic_loss}
\end{equation} 

This formulation ensures gradient flow throughout the entire pipeline, enabling the Visual Encoder (see Sec. \ref{learanbale}) to learn discriminative features for homophene disambiguation. The synergistic optimization of structural flow ($\mathcal{L}_{\mathrm{FM}}$), lexical precision ($\mathcal{L}_{\mathrm{CE}}$), and semantic consistency ($\mathcal{L}_{\mathrm{sem}}$) empowers our framework to robustly disambiguate homophenes.


\noindent\textbf{Adaptive Termination.} To enable the fixed-dimensional flow model to generate variable-length sequences, we introduce a Self-Terminating Decoder. We explicitly train the model to recognize sequence boundaries by stochastically appending padding tokens ($\mathtt{[PAD]}$) to the training targets \cite{ghazvininejad2019mask}. This compels the model to internalize the termination signal within the semantic manifold. The final transcription is truncated at the first termination token, enabling flexible sequence lengths.

\begin{figure}[t]
    \centering
    \includegraphics[width=1\linewidth]{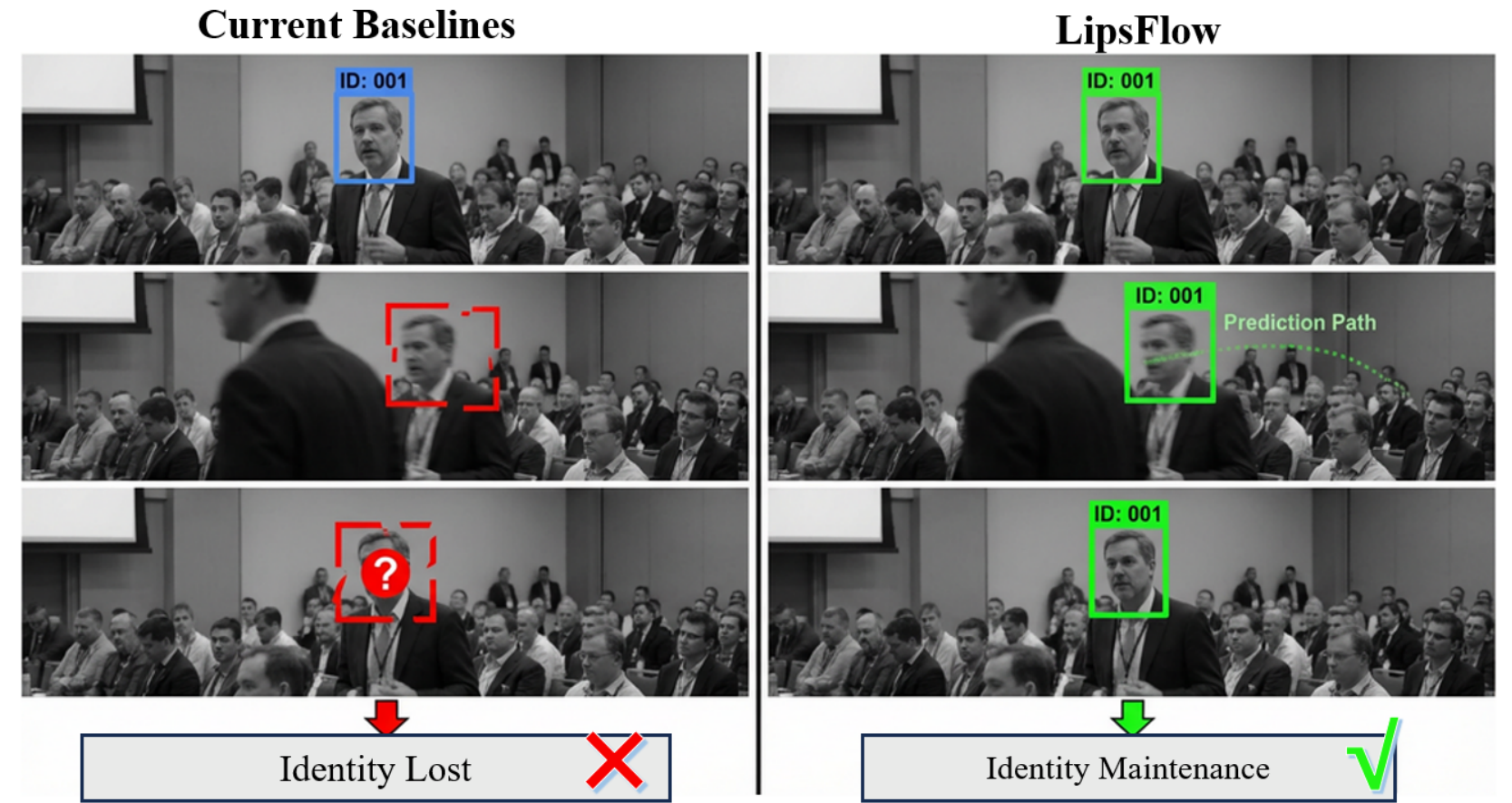}
    \caption{\textbf{Robustness of Identity Tracking in Multi-Speaker.} Existing baselines (\textbf{Left}) suffer from identity loss due to dynamic disturbances (\textcolor{red}{red} dashed boxes). In contrast, our method LipsFlow (\textbf{Right}) effectively maintains speaker identity continuity through its predicted path mechanism, ensuring stable identity tracking even in challenging environments (\textcolor{green}{green} solid boxes).}
    \label{pre-a}
\end{figure}

\section{Multi-Stage Data Processing Pipeline }
\label{data_process}

While Sec.  \ref{framework} details the LipsFlow architecture, its performance hinges on high-quality input. We establish a powerful \textbf{Multi-Stage Data Process Pipeline} designed to isolate target speakers in complex, multi-speaker environments. The pipeline implements a rigorous curation protocol to address three critical challenges: maintaining identity consistency under occlusion, extracting stable facial features, and filtering non-speech visual noise with specific annotation labels.

\subsection{Identity Consistency Maintenance}
\label{4.1}

We operate on the source RGB videos and build our data processing pipeline upon the robust ByteTrack~\cite{Zhang22-ByteTrack}. Specifically, we leverage its hierarchical recovery strategy supported by Kalman filtering \cite{Cao23-OCSORT} and IoU-based association \cite{Du23-StrongSORT} to rescue low-confidence detections from  \textbf{temporary occlusion} or \textbf{pose variations} (see Fig. \ref{pre-a}). 
Beyond applying standard tracking, we propose a tailored spatiotemporal trajectory refinement protocol optimized for event-based vision. We enforce a customized temporal continuity check~\cite{Qin24-MotionTrack} to actively merge fragmented segments and suppress transient false alarms in dynamic conversational scenarios. Significantly, we design this specialized pipeline not only to guarantee strictly aligned identity supervision without identity loss, but also to act as a dedicated physical noise-suppressor. 

\begin{figure*}[t]
    \centering
    \includegraphics[width=1\linewidth]{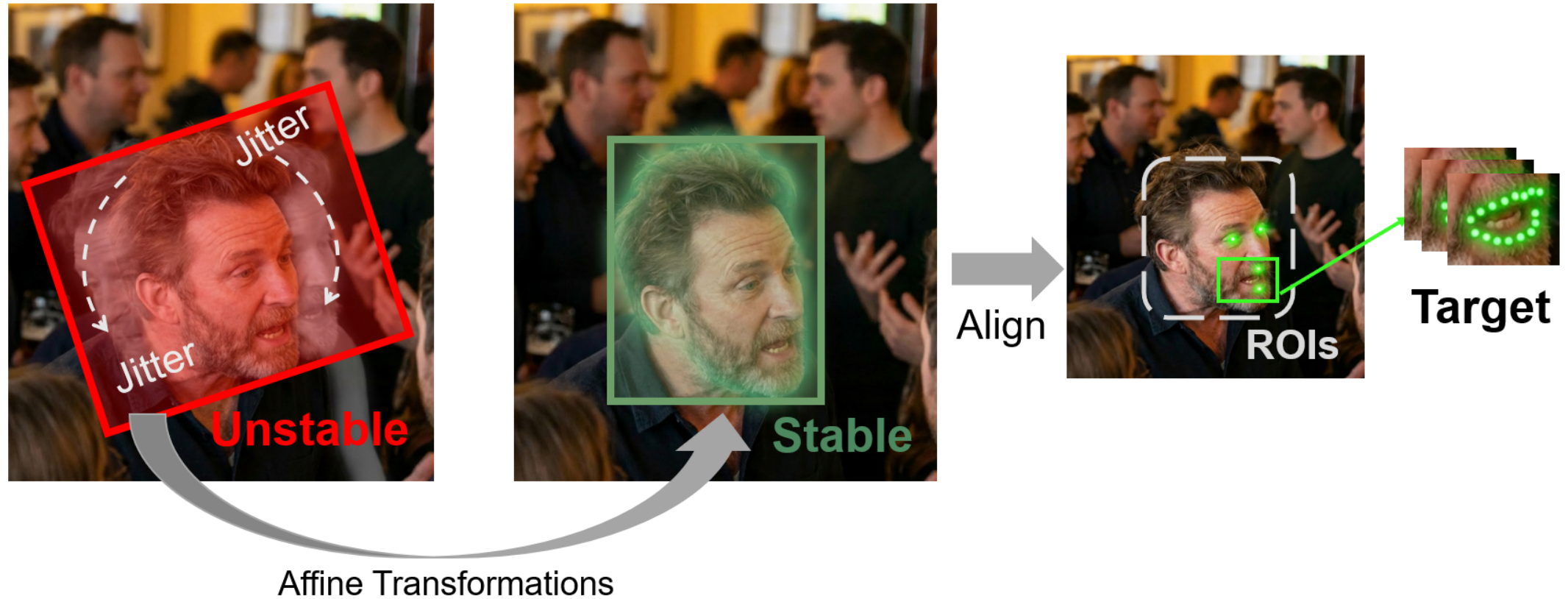}
    
    \caption{\textbf{Stabilization and Extraction of Features.} The video frames (\textcolor{red}{\textbf{Unstable}}) typically contain significant jitter, which leads to severe noise during events generation. By applying the Affine Transformation based on facial feature points, we align the ROIs to obtain a stable target region (\textcolor{green}{\textbf{Stable}}). This process ensures that subsequently generated events reflect only lip dynamics, not head movements.}
    \label{pre-b}
\end{figure*}

\subsection{ROIs Extraction and Stabilization}
\label{4.2}

We engineer a cascaded system for precise Lip Region of Interest (ROIs) extraction (see Fig.  \ref{pre-b}).
Following identity tracking, we deploy RetinaFace~\cite{deng2020retinaface} for high-precision face detection, succeeded by a Dlib~\cite{King09-Dlib} regression tree model to localize 68 facial landmarks.
We specifically extract the 20 lip landmarks (indices 48--67) incorporating a $20\%$ contextual padding. To eliminate rigid \textbf{head jitter} that corrupts event generation, we apply Affine Transformations~\cite{Ma23-AutoAVSR} to align the ROIs based on stable eye and nose anchors. To prevent false event triggers caused by artificial zero-padding boundaries introduced during this affine alignment, we apply a strict central crop and masking strategy. Finally, we normalize the purified crops to $128 \times 64$ using gradient-preserving interpolation, establishing a consistent, noise-free spatial reference tailored for downstream neuromorphic event streams generation with Sec. \ref{learanbale}.

\subsection{Annotation Protocol}
\label{annotation}

We establish an annotation protocol to distinguish valid speech sources from background noise. Following the common practice \cite{Roth20-AVA}, we define a three-tier vocabulary: (1) \textit{Silence}; (2) \textit{Oral Motion} (e.g., only chewing); and (3) \textit{Active}.
Fundamentally, this explicit state decoupling is an indispensable requirement for event-based VSR. Because neuromorphic sensors respond purely to temporal kinematics, state (2) triggers dense event streams that are visually indistinguishable from actual speech, which would otherwise induce severe semantic hallucinations in generative models. To resolve these visual-audio discrepancies, we leverage synchronized audio waveforms as a semantic ground-truth anchor to precisely calibrate the temporal boundaries of (3). 
\begin{figure}[htbp]
    \centering
    \includegraphics[width=1\linewidth]{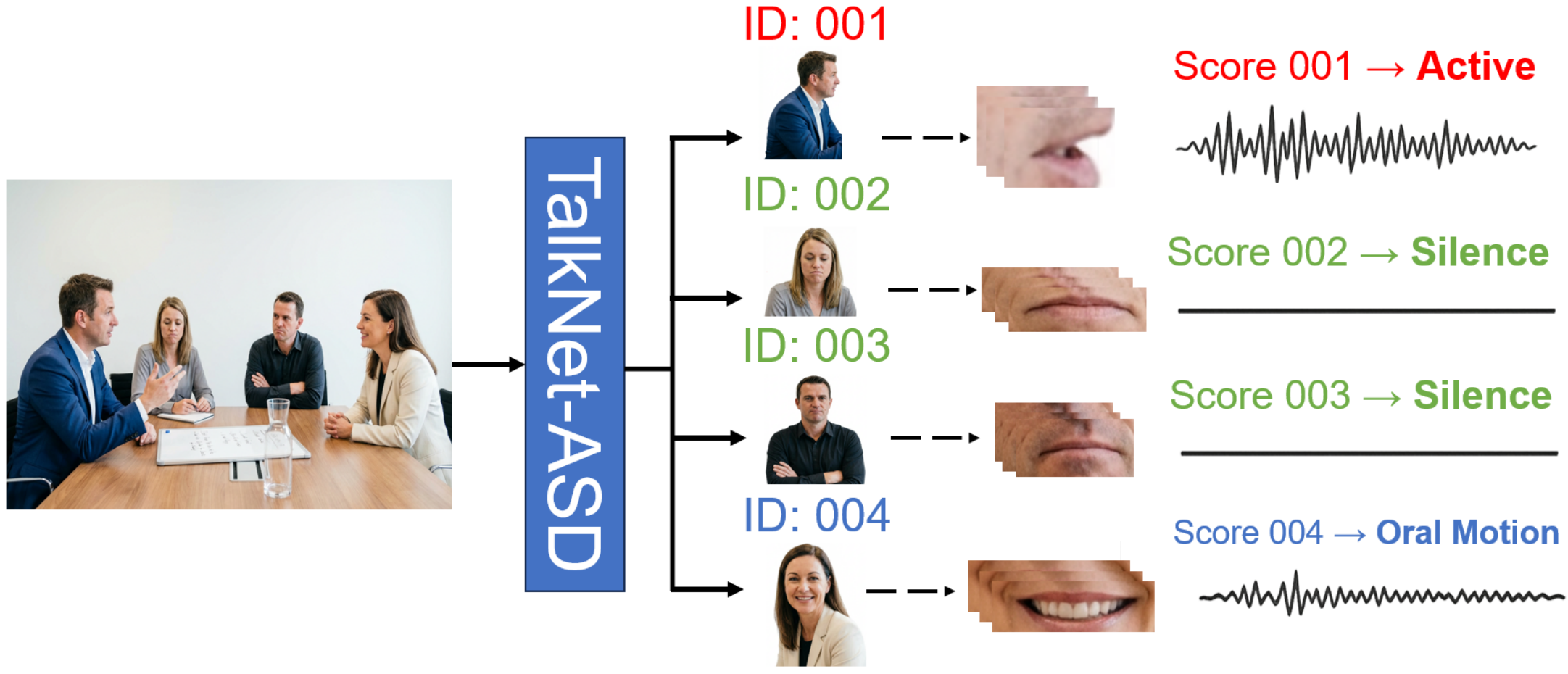}
    \caption{\textbf{Multi-Speaker Interference with Active Speaker Detection.} Our system assigns a specific annotation state to the tracked identity (see Sec. \ref{annotation}), enabling model to pinpoint the primary speaker and ignore visual interference from passive listeners. Note, we retain only identity-audio pairs exceeding a strict confidence threshold.}
  
    \label{pre-c}
    \end{figure}

\subsection{Multi-Speaker Active Detection}
\label{4.4}
While ByteTrack~\cite{Zhang22-ByteTrack} maintains visual identity, the association of these identities with overlapping audio streams in multi-speaker scenarios presents a distinct challenge.
We resolve this \textbf{identity-to-audio} mapping problem by leveraging TalkNet-ASD~\cite{Tao21-TalkNet} to improve the accuracy of audio-visual consistency in matching identity. As illustrated in Fig.  \ref{pre-c}, this strategy computes correspondence scores between each tracked identity and temporal audio segments by evaluating the correlation between macroscopic lip motions and mel-spectrogram features.

\section{Experiments and Results}

\subsection{Implementation Details}
\label{implement}

\textbf{Datasets.} We evaluate LipsFlow and baselines on two complementary benchmarks: \textbf{DVS-Lip}\cite{Tan22-MSTP} and \textbf{AVA}\cite{Roth20-AVA}. The former constitutes the first large-scale corpus with real-world event streams, encompassing not only subtle lip micro-movements but also challenging vocabulary. Note, we utilize visual information (voxel or event streams) directly from DVS-Lip. Conversely, the AVA introduces uncontrolled, natural scene complexity multi-speaker scenarios, which can convert frames into event streams via our Data Process Pipeline (see Sec. \ref{data_process}). We perform three classification annotations on the AVA with video keyword retrieval and human screening, categorizing it into \textit{Rapid Motion}, \textit{Low Light}, and \textit{Severe Occlusion}, to validate the model's adaptability for complex scenarios.

\noindent \textbf{Data Augmentation.} We further validate the adaptability of LipsFlow and baselines to the environment through a standard data augmentation mechanism. This process includes four probabilistic transformations: \textbf{MixUp} (probability 0.3)~\cite{Zhang18-Mixup}, which promotes linear behavior between categories through Beta distribution interpolation; \textbf{CutMix} (probability 0.3)~\cite{Yun19-CutMix}, which replaces fragments with consecutive 25\% to 50\% segments, forcing the network to learn temporal continuity rather than memorizing local patterns; \textbf{Occlusion} (probability 0.2), which simulates sensor malfunctions or frame jitter by masking subsequences~\cite{Wang23-VideoMAE_V2}; and \textbf{Identity Mapping} (probability 0.2), used to maintain the fidelity of the distribution. This comprehensive strategy effectively simulates imperfections in real-world videos, thus ensuring the model's robustness~\cite{Ma23-AutoAVSR}.
\label{data_aug}

\begin{table*}[t]
\scriptsize
\centering
\caption{Comparison Benchmarking Results on \textbf{DVS-Lip}. Note that  we utilize the event streams provided by this dataset directly as input, rather than simulations. Metrics: \textit{WER} \textit{/}\textit{VER}  (Word \cite{Assael2016LipNet}/Viseme \cite{Tan22-MSTP} Error Rate, \%, $\downarrow$), \textit{RTF}  (Real-Time Factor \cite{ren2021fastspeech}, s), \textit{Latency} (inference in ms), and \textit{Params} (Model Parameters, M). \textbf{Bold} and \underline{underlined} values denote the best and second-best performance. While Llama-AVSR \cite{Pan24-LLMVSR} achieves competitive accuracy, our method balances high fidelity with superior efficiency. The principles for reproduction all follow those outlined on the official website. }
\label{tab:main_results}

\setlength{\tabcolsep}{0pt} 

\begin{tabular*}{\linewidth}{@{\extracolsep{\fill}} l c c cc ccc }
\toprule

\multirow{2}{*}{\textbf{Method}} & 
\multirow{2}{*}{\textbf{Modality}} & 
\multirow{2}{*}{\textbf{Backbone}} & 
\multicolumn{2}{c}{\textbf{Accuracy ($\downarrow$)}} & 
\multicolumn{2}{c}{\textbf{Efficiency ($\downarrow$)}} \\ 
\cmidrule{4-5} \cmidrule{6-7}

 & & & \textbf{WER}& \textbf{VER}& \textbf{RTF} & \textbf{Latency} & \textbf{Params} \\ 
\midrule

\multicolumn{8}{l}{\textit{Video-based: Standard Architectures}} \\
Auto-AVSR~\cite{Ma23-AutoAVSR} & RGB & Conformer & 28.4 & 26.1 & 0.45 & 580 & 256.4 \\
VATLM~\cite{Zhu23-VATLM}       & RGB & Transformer   & 26.9 & 25.3 & 0.52 & 650 & 310.8 \\
RAVEn~\cite{Haliassos23-RAVEn} & RGB & Conformer   & 31.2 & 29.8 & \textbf{0.15} & \textbf{200} & 55.2 \\ 
SynthVSR~\cite{Liu2023SynthVSR}    & RGB & Conformer & 25.1 & 23.5 & 4.80 & 6000 & 185.6 \\ 

\midrule

\multicolumn{8}{l}{\textit{Video-based: Recent SOTA}} \\
CMAVSR~\cite{Shi24-CMVSR}      & RGB & Transformer    & 27.5 & 25.8 & 0.46 & 590 & 260.1 \\
LipGen~\cite{Hao25-LipGen}            & RGB & ResNet-18 & 26.1 & 24.6 & 0.55 & 680 & 235.5 \\
Llama-AVSR~\cite{Pan24-LLMVSR}    & RGB & LLM    & \underline{23.9} & \underline{21.5} & 2.10 & 2500 & $>$1000 \\

\midrule

\multicolumn{8}{l}{\textit{Event-based Methods}} \\
MSTP~\cite{Tan22-MSTP}        & Event & ResNet-18        & 30.5 & 28.2 & 0.22 & 280 & 22.5 \\
SNN-Lip~\cite{bulzomi2023end}   & Event & SNNs     & 34.1 & 32.4 & \underline{0.18} & 230 & \textbf{12.1} \\ 
\midrule

\rowcolor{gray!25} 
\textbf{LipsFlow (Ours)} & Event & \textbf{CFM-Former} & \textbf{22.3} & \textbf{19.8} & \underline{0.18} & \underline{240} & \underline{45.8} \\ 
\bottomrule
\end{tabular*}

\end{table*}

\noindent\textbf{Training.} We orchestrate the training process on 4 NVIDIA A100 (80GB) GPUs via the PyTorch Distributed Data Parallel framework. We strictly follow the official train/test protocols of the DVS-Lip \cite{Tan22-MSTP} and AVA \cite{Roth20-AVA} datasets to ensure a fair comparison with baselines. Furthermore, we leverage BF16 mixed precision~\cite{Kalamkar19-BF16} and TF32 acceleration~\cite{Choquette21-A100} under a global batch size of 64 (32 sequences per GPU) to balance numerical fidelity with computational throughput. We design a rigorous two-stage curriculum to decouple feature extraction from generation: initially, we \textit{freeze} the visual front-end for the first 10 epochs (\textbf{$\approx$ }15k iterations, $lr = 5 \times 10^{-4}$) to align the generative manifold without destabilizing the pretrained representations; subsequently, we \textit{unfreeze} the entire architecture for holistic end-to-end fine-tuning for another 20 epochs (\textbf{$\approx$ }30k iterations, $lr = 1 \times 10^{-4}$). We drive convergence using the AdamW \cite{Loshchilov19-AdamW} optimizer ($\beta_1=0.9, \beta_2=0.995$) and safeguard training stability via gradient norm clipping (1.0). The learning rate is modulated by a cosine annealing scheduler with 2,000 warmup steps, concluding the entire 30 hours training regimen after approximately 45k total iterations.

\begin{table}[t]
\centering
\scriptsize
\caption{Robustness Benchmarking Results on \textbf{AVA} with WER (\%, $\downarrow$). \textit{Clean Set} aligns with the main results in Tab. \ref{tab:main_results}. Values in parentheses denote relative degradation ($\uparrow$ performance drop). Values in red/orange/green font indicate severe/moderate/slight performance degradation, respectively, with smaller values representing better results. Note, we use the  adaptive threshold network (see Sec. \ref{learanbale}) and data processing pipeline (see Sec. \ref{data_process}) to generate the event streams.   }
\label{tab:robustness}

\setlength{\tabcolsep}{5pt}

\begin{tabular}{l|c|ccc}
\toprule
\multirow{2}{*}{\textbf{Method}} & \textbf{Clean} & \textbf{Rapid} & \textbf{Low} & \textbf{Severe} \\
& \textbf{Set} & \textbf{Motion} & \textbf{Light} & \textbf{Occlusion} \\ \midrule

\textit{Video-based} & & & & \\
Auto-AVSR~\cite{Ma23-AutoAVSR} & 28.4 & 45.2 {\color{red}($\uparrow$16.8)} & 52.1 {\color{red}($\uparrow$23.7)} & 48.9 {\color{red}($\uparrow$20.5)} \\
LipGen~\cite{Hao25-LipGen} & 26.1 & 41.8 {\color{red}($\uparrow$15.7)} & 49.5 {\color{red}($\uparrow$23.4)} & 44.2 {\color{red}($\uparrow$18.1)} \\
SynthVSR~\cite{Liu2023SynthVSR} & 25.1 & 39.4 {\color{red}($\uparrow$14.3)} & 47.1 {\color{red}($\uparrow$22.0)} & 38.6 {\color{red}($\uparrow$13.5)} \\
Llama-AVSR~\cite{Pan24-LLMVSR} & 23.9 & 36.5 {\color{red}($\uparrow$12.6)} & 44.3 {\color{red}($\uparrow$20.4)} & 35.1 {\color{orange}($\uparrow$11.2)} \\ \midrule

\textit{Event-based} & & & & \\
MSTP~\cite{Tan22-MSTP} & 30.5 & 33.1 {\color{green}($\uparrow$2.6)} & 31.5 {\color{green}($\uparrow$1.0)} & 42.8 {\color{orange}($\uparrow$12.3)} \\ \midrule

\rowcolor{gray!25}
\textbf{LipsFlow (Ours)} & \textbf{22.3} & \textbf{24.1} {\color{green}($\boldsymbol{\uparrow}$1.8)} & \textbf{22.9} {\color{green}($\boldsymbol{\uparrow}$0.6)} & \textbf{26.5} {\color{green}($\boldsymbol{\uparrow}$4.2)} \\ \bottomrule
\end{tabular}
\end{table}

\noindent\textbf{Baseline Systems.} We compare LipsFlow within selective landscapes across two modalities (RGB and Events). Regarding \textbf{Video-based} methods, we benchmark against the established \textit{Auto-AVSR}~\cite{Ma23-AutoAVSR} and the pre-trained \textit{VATLM}~\cite{Zhu23-VATLM} with Masked Prediction architecture, while incorporating current SOTA research: the robust \textit{RAVEn}~\cite{Haliassos23-RAVEn}, the contrastive \textit{CMAVSR}~\cite{Shi24-CMVSR}, the generation-enhanced \textit{LipGen}~\cite{Hao25-LipGen}, and the large-model-augmented \textit{Llama-AVSR}~\cite{Pan24-LLMVSR}. Furthermore, we juxtapose our generative efficiency with the Conformer-based \textit{SynthVSR}~\cite{Liu2023SynthVSR}, highlighting our OT-CFM paradigm. We scrutinize our performance against specialized \textbf{Event-base}d methods, specifically the \textit{MSTP}~\cite{Tan22-MSTP} and the parameter-efficient \textit{SNN-Lip}~\cite{bulzomi2023end}, to maintain the comprehensiveness of the RGB and Events baselines comparison. Note, the video-based methods input only uses RGB video information, while the event-based methods input requires additional generated neuromorphic event streams for a fair comparison.

\subsection{Comparative Benchmarking Results}
\label{tab1_analysis}

As quantified in Tab. \ref{tab:main_results}, we demonstrate that LipsFlow establishes a new performance ceiling on the multi-speaker event-based benchmark DVS-Lip for VSR, securing a WER of \textbf{22.3\%} and a VER of \textbf{19.8\%}. 
Notably, our method outperforms the sophisticated video-based baselines, surpassing the generation-enhanced LipGen by 3.8\% and even edging out the large-scale Llama-AVSR by 1.6\%. This result highlights a critical insight: our method captures transient phonetic micro-dynamics, such as rapid plosives, more effectively than simply scaling up model parameters or utilizing massive pre-trained language backbones on standard RGB video. 
Simultaneously, we surpass the leading event-based competitor (MSTP) by 8.2\%, validating the architectural superiority of our CFM-Former backbone over traditional ResNet-18 paradigms in modeling complex temporal dependencies. 
While our WER (22.3\%) is highly competitive with the Conformer-based SynthVSR (22.5\%), we drastically slash the RTF from 4.80 to 0.18, a staggering 26× acceleration. This speedup stems directly from our introduced OT path, which enables high speed inference in just 2 Number of Function Evaluations (NFE) compared to prior methods. 

\begin{table}[t] 
\centering
\scriptsize 
\setlength{\tabcolsep}{2.5pt} 
\caption{\textbf{Ablation Study on Architecture Variants.} 
\textit{Fusion Strategy} contrasts naive channel-wise concatenation (\textit{Concat}) with our dual-branch temporal alignment via Cross-Attention (\textit{Cross-Attn}).
Paradigm benchmarks standard Auto-Regressive (\textit{AR})  Transformers and Denoising Diffusion (\textit{DDPM}) against our OT-CFM decoder.}
\label{tab:ablation_component}

\begin{tabular}{l c c c c c c c}
\toprule
\multirow{2}{*}{\textbf{Variants}}& \multicolumn{2}{c}{\textbf{Input}} & \textbf{Fusion} & \textbf{Decoder} & \multicolumn{2}{c}{\textbf{Accuracy ($\downarrow$)}} & \\
\cmidrule(lr){2-3} \cmidrule(lr){6-7}
 & \textbf{RGB} & \textbf{Event} & \textbf{Strategy} & \textbf{Paradigm} & \textbf{WER} & \textbf{VER} & \textbf{Params} \\ 
\midrule
(a) Visual-Only& \checkmark & \ding{55} & N/A & AR & 27.8 & 25.9 & 42.1 \\
(b) Naive Fusion & \checkmark & \checkmark & Concat & AR & 25.9 & 24.1 & 44.5 \\
(c) Neuromorphic Branch& \checkmark & \checkmark & Cross-Attn & AR & 24.5 & 22.8 & 46.2 \\
(d) Diffusion Variant & \checkmark & \checkmark & Cross-Attn & Diffusion & 22.5 & 20.1 & 185.6 \\
\midrule
\rowcolor{gray!25} 
\textbf{(e) LipsFlow (Ours)} & \checkmark & \checkmark & \textbf{Cross-Attn} & \textbf{OT-CFM} & \textbf{22.3} & \textbf{19.8} & \textbf{45.8} \\ 
\bottomrule
\end{tabular}
\end{table}

\subsection{Environmental Resilience and Degradation}
\label{rob_analysis}

As quantified in Tab. \ref{tab:robustness}, LipsFlow exhibits exceptional resilience against environmental degradation. A pivotal finding is that the limitations of RGB-based sensing persist regardless of algorithmic sophistication. Even the generative model SynthVSR and the large-scale LLM-augmented systems Llama-AVSR suffer catastrophic performance drops. This stems from the fact that pure-pixel sensors are physically incapable of recovering articulatory details lost to motion blur and photon scarcity. In contrast, by shifting the sensing paradigm to neuromorphic event streams, LipsFlow maintains near-negligible degradation (e.g., only 
↑0.6\% in Low Light).  Furthermore, while Llama-AVSR leverages massive linguistic priors to mitigate Severe Occlusion, our method still outperforms it by a significant margin (26.5\% vs. 35.1\% WER), demonstrating that neuromorphic event streams representation provides more effective and richer visual information for VSR compared to traditional video-based recognition methods.

\subsection{Ablation Study}

We conduct a comprehensive ablation study on the \textbf{DVS-Lip} \cite{Tan22-MSTP} to verify the effectiveness of our method components. We specifically ablate our multi-modal fusion, OT-CFM decoder efficiency, and joint optimization objectives.\\

\noindent\textbf{Impact of Multi-modal Fusion and Architecture.} Tab. \ref{tab:ablation_component} confirms that the Visual-Only (a) variant achieves a WER of 27.8\%, verifying the physical bottleneck of RGB sensors in capturing transient phonemes under rapid articulatory motion. Introducing the Neuromorphic Branch (c) via our Cross-Attention fusion breaks this ceiling to 24.5\% by effectively aligning asynchronous event spikes with synchronous video frames, significantly outperforming the channel-wise naive concatenation strategy in (b). Furthermore, replacing standard decoders with the generative LipsFlow (e) yields the final leap; while the Diffusion variant (d) is accurate, it is parameter-heavy (185.6M), whereas our full model achieves the best WER (22.3\%) with 4$\times$ fewer parameters, demonstrating superior efficiency in modeling complex phonetic distributions within the semantic latent space.\\

\begin{table}[t]
\centering
\scriptsize
\caption{\textbf{Ablation Study on Inference Dynamics.} \textit{NFE} vs. \textit{Efficiency} Trade-off.
Analyzing sampling steps against fidelity and speed. 
\textit{Speedup}: Relative to Diffusion. }
\label{tab:ablation_nfe}
\setlength{\tabcolsep}{6pt}
\begin{tabular}{l|c|c|ccc}
\toprule
\textbf{Variants}& \textbf{NFE} & \textbf{WER$\downarrow$}& \textbf{Latency$\downarrow$}& \textbf{RTF$\downarrow$} & \textbf{Speedup$\uparrow$} \\
\midrule
 & 1 & 28.5 & 180 & 0.09 & 33.3$\times$ \\
\rowcolor{gray!25} & \textbf{2} & \textbf{22.3} & \textbf{240} & \textbf{0.18} & \textbf{25.0$\times$} \\
 & 4 & 22.1 & 350 & 0.26 & 17.1$\times$ \\
\multirow{-5}{*}{\textbf{LipsFlow (Ours)}}& 10 & 22.0 & 680 & 0.51 & 8.8$\times$ \\ \midrule
\multicolumn{6}{l}{\textit{Comparison Results:}} \\
Diffusion Variant & 50 & 22.5 & 6000 & 4.80 & 1.0$\times$ \\
\bottomrule
\end{tabular}

\end{table}

\begin{table}[t]
\centering
\scriptsize
\caption{\textbf{Ablation Study on Loss Configuration and Training Strategy.} \textit{CE} represents the Cross-Entropy loss. We verify the stepwise contribution of optimization objectives and the 2-stage curriculum.
\textit{SpkVA}: Speaker Verification Accuracy \cite{Chung18-VoxCeleb2}.}
\label{tab:ablation_strategy}

\setlength{\tabcolsep}{6pt}

\begin{tabular}{l|c|cc}
\toprule
\textbf{Loss Configuration} & \textbf{Stage} & \textbf{WER$\downarrow$} & \textbf{SpkVA$\uparrow$}\\
\midrule
CE-Only& Single & 35.8 & 76.3 \\
OT-CFM& Single & 28.2 & 78.5 \\
+ Speaker Contrastive & Single & 26.6 & 91.7 \\
Full Loss& Single& 25.2& 92.3\\ 
\midrule
\multicolumn{4}{l}{\textit{Impact of Training Strategy (with Full Loss):}} \\
Full Loss & Single & 25.2 & 92.3 \\
Full Loss& 2-Stage& 23.8& 93.1\\
\rowcolor{gray!25} Full Loss + Data Aug& \textbf{2-Stage}& \textbf{22.3} & \textbf{93.8} \\
\bottomrule
\end{tabular}
\end{table}

\noindent\textbf{Efficiency of Rectified Flow Trajectories.} Tab. \ref{tab:ablation_nfe} shows the critical speed-accuracy trade-off.While single-step generation of NFE suffers from a notable truncation error (28.5\% WER) due to the slight residual curvature in the learned vector field, increasing the budget to just NFE=2 yields an optimal performance of 22.3\%. This dramatic 6.2\% improvement with only one additional integration step empirically validates the core advantage of our OT objective: it geometrically straightens the probability flow so effectively that a simple 2-step Euler solver can traverse the generative manifold with almost zero discretization error. Consequently, we achieve an RTF of 0.18, a massive 25.0$\times$ speedup over the curved trajectory of Diffusion (6000ms), thereby ensuring real-time feasibility.\\

\noindent\textbf{Optimization and Training Strategy.} Tab. \ref{tab:ablation_strategy} illustrates the systematic contribution of each learning component. Replacing the discriminative CE-Only (35.8\% WER) with our generative OT-CFM drastically lowers WER to 28.2\%, proving that modeling \textbf{continuous} probability flows \textbf{along optimal trajectories} better captures phonetic diversity than simple classification. Additionally, we introduce Speaker Contrastive Learning to decouple features, improving SpkVA to 92.3\% by effectively filtering out identity-irrelevant visual disturbances. Regarding the curriculum, our proposed 2-Stage Strategy (freeze-then-fine-tune) prevents the collapse of pre-trained manifolds during initial convergence, yielding a 1.4\% gain. Finally, applying the Data Augmentation (\textit{Data Aug}) in Sec. \ref{implement} (MixUp, CutMix) secures the performance WER of 22.3\%, ensuring robust generalization against unforeseen real-world noise and environmental variability.

\section{Conclusion}

In this work, we first present \textbf{LipsFlow}, a novel framework that fundamentally reimagines VSR by bridging neuromorphic perception with deterministic generative modeling. To overcome the physical limitations of standard cameras and the prohibitive latency of diffusion models, we introduce a paradigm built on two core pillars. We leverage high-frequency event streams to capture transient lip dynamics, completely bypassing the redundancy and motion blur of frame-based video. We employ OT-CFM to learn deterministic, straight-line ODE trajectories, thereby resolving the historical tension between generation quality and inference speed with our data processing pipeline. Furthermore, we address the inherent challenge of homophene ambiguity and  non-phonetic oral recognition, which is precisely what traditional methods have failed to solve.

\section*{Acknowledgements}

This work was supported in part by National Natural Science Foundation of China (No. 62577004) and Beijing Natural Science Foundation (No. L2611045).

%
%
\bibliographystyle{splncs04}
\bibliography{main}
\end{document}